\documentclass{aa}  
\usepackage{rotating}
\usepackage{natbib}
\usepackage{amsmath}
\bibpunct{(}{)}{;}{a}{}{,} 
\usepackage{float}
\usepackage{placeins}

\newcommand{\kms}{km s$^{-1}$}

\newcommand{\bp}{$\beta$\,Pic}

\usepackage{color}
\usepackage{hyperref}
\hypersetup{colorlinks=true,allcolors=[rgb]{0,0,0.8}}

\usepackage[nolist,nohyperlinks]{acronym}
\newacro{harps}[HARPS]{High Accuracy Radial Velocity Planet Searcher}
\newacro{uves}[UVES]{Ultra-violet Visible Echelle Spectrograph}
\newacro{feb}[FEBs]{Exocomets}

\usepackage{graphicx}
\usepackage{txfonts}

\usepackage{showyourwork}

\usepackage[version=4]{mhchem}
\makeatletter
\renewcommand*\aa@pageof{, page \thepage{} of \pageref*{LastPage}}
\makeatother

\begin{document} 
   \title{Upper limits on CN from exocomets transiting $\beta$~Pictoris}
   \author{M.A. Kenworthy
          \inst{1}
          \and
          E. de Mooij\inst{2}
          \and
          A.\ Brandeker\inst{3}
          \and 
         C. Opitom\inst{4}
          \and
          F. Kiefer\inst{5}
          \and
          A. Fitzsimmons \inst{2}
          }
   \institute{Leiden Observatory, Leiden University, P.O. Box 9513, 2300 RA Leiden, The Netherlands\\
              \email{kenworthy@strw.leidenuniv.nl}
        \and
        Astrophysics Research Centre, School of Mathematics and Physics, Queen's University Belfast, BT7 1NN Belfast, UK
        \and
        Institutionen f\"{o}r astronomi, Stockholms universitet, AlbaNova universitetscentrum, 106 91, Stockholm, Sweden
        \and
        Institute for Astronomy, University of Edinburgh, Royal Observatory, Edinburgh EH9 3HJ, UK
        \and
        LESIA, Observatoire de Paris, Universit\'{e} PSL, CNRS, 5 Place Jules Janssen, 92190 Meudon, France}

   \date{Received 7 Feb 2025; Accepted 13 Apr 2025}

  \abstract
   {The young (23 Myr) nearby (19.4 pc) star $\beta$~Pictoris hosts an edge-on debris disk with two gas giant exoplanets in orbit around it.
   Many transient absorption features have been detected in the rotationally broadened stellar lines, which are thought to be the coma of infalling exocomets crossing the line of sight towards Earth.}
   {In the Solar System, the molecule cynaogen (\ce{CN}) and its associated ionic species are one of the most detectable molecules in the coma and tails of comets.
   We perform a search for cyanogen in the spectra of \bp{} to detect or put an upper limit on this molecule's presence in a young, highly active planetary system.}
   {We divide twenty year's worth of \ac{harps} spectra into those with strong exocomet absorption features, and those with only stellar lines.
   The high signal-to-noise stellar spectrum normalises out the stellar lines in the exocomet spectra, which are then shifted and stacked on the deepest exocomet absorption features to produce a high signal-to-noise exocomet spectrum, and search for the \ce{CN} band head using a model temperature dependent cross-correlation template.}
   {We do not detect \ce{CN} in our data, and place a temperature and broadening dependent 5$\sigma$ upper limit between 10$^{12}$\,cm$^{-2}$ and 10$^{13}$\,cm$^{-2}$, to be compared to the typical 10$^9$ - 10$^{10}$ cm$^{-2}$ expected from scaling of the values in the Solar System comets.}


   \maketitle

\section{Introduction}

The formation and evolution of planets and their attendant moons is thought to occur in the first several million years of a planetary system.
Within our Solar System, comets deliver volatiles from the outermost reaches of the Oort cloud \citep{Obrien2018}, past the gas giants and terrestrial planets, even as close as a few solar radii.
It is therefore important to understand the distribution, nature and chemical composition of exocomets in other planet forming systems.

One of the closest young planetary systems is $\beta$~Pictoris, which has been studied intensively both with direct imaging and with spectroscopy.
\bp{} is a young \citep[$\sim 25$ Myr; see ][ for discussions]{Lee24}, nearby \citep[$19.63\pm 0.06$ pc, ][]{Lindegren21} and early type star \citep[A6V; ][]{Gray06} which hosts an edge-on debris disk and at least two exoplanets, making it one of the most intensely studied young planetary systems.
The star is a rapid rotator \citep[$124\pm 3$\,\kms{}; ][]{Koen03} seen close to equator on.
Amongst these, the \ce{Ca} H and K lines show deep absorption down to 0.1 of the continuum level.
Deep within this rotationally broadened line there is a consistent and stable narrow absorption feature with a full width half maximum (FWHM) of a few \kms{}, attributed to the presence of the circumstellar material \citep{Hobbs85,VidalMadjar86}.

Transient absorption features appearing in the rotationally broadened \ce{Ca} H and K features were proposed to be due to the transit of evaporating exocomets in front of the star \citep{Ferlet87}.
These features appear and disappear on the timescale of hours to days, with some features lasting up to tens of days and have red-shifted velocities from eighty kilometres per second, all the way down to (and merging with) the circumstellar line, with a few appearing on the blueward side \citep{LagrangeHenri92} with widths from a few up to tens of kilometres per second.
The absorption can be described by the transit of evaporating bodies that pass within a few stellar radii of the star, resulting in sublimation of material that produce a large coma containing calcium ions which are then seen as an additional absorption component \citep{Beust91}.
Intensive spectroscopic monitoring over several hours show both multiple components transiting the disk of the star \citep{Beust96} and changes of radial velocity of several components are consistent with Keplerian elliptical orbits \citep{Kennedy18} implying at least two \citep{Kiefer14,Pollard24} or more \citep{Heller24} distinct kinematic families of Beta Pictoris exocomets.
At least one event is derived with the orbit parallel to the rotational axis of the star \citep{Tobin19}.

From the multiple detections of exocomets in spectroscopy towards \bp{}, the broadband detection of exocomet tails transiting a star were predicted in \citet{1999A&A...343..916L}.
A chain of seven exocomets were proposed to explain the shape of one eclipse seen towards KIC~8462852 in \citet{Kiefer17}, and
single exocomet transits were subsequently detected towards the star KIC~3542116 \citep{Rappaport18} with the {\it Kepler} satellite \citep{Borucki10} and in TESS data towards \bp{} \citep{Zieba19}.
Observations of \bp{} in subsequent TESS sectors revealed over two dozen transits whose depth follows a power law distribution similar to those of the Solar System \citep{LecavelierdesEtangs22,Pavlenko22}.
Automated searches in photometry \citep{Kennedy19} and spectroscopy \citep{BendahanWest24} for additional exocomet systems have been met with limited success.

To the contrary of what is observed in the case of Beta Pictoris exocomets, most molecular and atomic signatures detected in the coma of Solar System comets are emission lines (or bands) instead of absorptions.
Spectroscopic observations of Solar System comets when they are very close to the Sun are rare due to observational difficulties.
The only comet spectroscopically observed closer than 0.2~au from the Sun is C/1965~S1 Ikeya-Seki.
For this comet, it was noticed that the main emission bands or lines detected in its optical spectrum dramatically changed while the comets approached the Sun.
At distances greater than 0.4~au, \ce{CH}, \ce{CN}, \ce{C2}, and \ce{C3} emission bands dominated the spectrum.
However, at 0.15~au from the Sun, the only molecular emission detected was \ce{CN}, while numerous atomic lines were visible \citep[\ion{[O]}{i}, \ion{Na}{i}, \ion{K}{i}, \ion{Ca}{ii}, \ion{Cr}{i}, \ion{Mn}{i}, \ion{Fe}{i}, \ion{Ni}{i}, \ion{Cu}{i}, \ion{V}{i}; ][]{Dufay65,Thackeray1966,Preston1967,Slaughter1969}.
Additionally, a tail of \ion{Fe}{i} atoms was proposed to explain a linear feature imaged in the tail of sungrazing comet C/2016~P1 McNaught 2006P1 \citep{Fulle2007} and \ce{Na} and \ce{K} were detected for comet C/2011 L4 \citep{Fulle13} at 0.46 au. 

On the other hand, \ion{Na}{i} D-line emission has been observed in many comets within $r \sim 1$ au of the Sun, where either the emission was much brighter than the foreground sky emission, or the comet was observed at high spectral resolution enabling separation from the telluric lines \citep{Cremonese02,Schmidt16}.
The high efficiency of the \ion{Na}{i} D transition makes it easily detectable, even for relatively low column density.
A notable discovery was a spectacular tail of \ion{Na}{i} associated with comet C/1995 O1 Hale-Bopp at $r\sim 1$ au, formed by radiation pressure on the neutral atoms before ionisation \citep{Cremonese1997}.
Recently, \cite{Hui2023} also demonstrated thet the activity of near-Earth asteroid Phaeton, with a perihelion of only 0.14 au, is associated with emission of sodium. 

The detection of these atoms only at small heliocentric distances imply that \ion{Ca}{ii} and \ion{Fe}{i} are predominantly locked up within cometary dust gains, and only observed when a comet is close enough to the Sun to allow their sublimation.
This occurs at $r\leq 0.03-0.06$ au for a dust grain with Bond albedo $\sim0.01$ and a sublimation temperature of $T\simeq 1600$K, depending on the thermal model assumed.
On the other hand, the origin of the \ion{Na}{i} tail observed in numerous comets at larger distances from the Sun is still not very well understood \citep{Cremonese02}.
Release from dust grains as well as molecular processes have been suggested as possible sources of \ion{Na}{i} in the coma and tail of comets annd explain its presence at large solar distances \citep{Cremonese1997}.

For the majority of Solar System comets, their optical spectra are dominated by the reflected sunlight from dust grains in their comae, plus molecular and atomic emission lines \citep[e.g. ][]{hyland2019}.
Most of the neutral and ionic molecular emission results from resonance fluorescence with solar photons.
Some species, such as \ce{H2O+} and \ce{CO+} result from ionization of parent molecules (i.e. molecules released directly by the sublimation of nuclear ices), all other molecules observed at optical wavelengths are second generation species, often created via photodissociation of parent molecules or released by organic-rich grains in the coma.
Another process also observed in optical spectrum of comets is prompt atomic emission  ([\ion{O}{i}], [\ion{C}{i}], and [\ion{N}{i}]).
Those forbidden transitions are the result of atoms produced in an excited state by the photodissociation of parent molecules in the coma. 

Of all cometary gas emission, the highest signal to noise normally occurs with the \ce{CN} (0-0) band at $\lambda\simeq 3880$\AA.
It is usually the first emission detected in the optical spectrum of a comet as it approaches the Sun, and as been detected in the comae of comets observed at relatively large distances from the Sun \citep{Bus1991,Fitzsimmons1996}.
The relative abundance in the coma of most comets of \ce{OH/CN}$\simeq500$ \citep{AHearn1995}, together with its high fluorescence scattering efficiency factor of approximately 2.6 photons per second per molecule at 1~au from the Sun and $\dot{r}\simeq 0$ \citep{Schleicher2010}, gives rise to a clear signature of a cometary gas coma.
Historically, the \ce{CN} was believed to be created from photodissociation of \ce{HCN}, which is present in abundances of the order of 1\% relative to water in cometary ices.
High-spatial resolution sub-mm mapping of \ce{HCN} distributions supports a nuclear source in some comets \citep{Cordiner2014}.
However, it is now known that in many comets a substantial fraction of \ce{CN} is released from sub-micron dust grains in the inner coma of comets \citep[e.g. ][]{Fray2005}. 

The \ce{C2} (0-0) band at $\lambda\simeq 5160$\AA{} may be of comparable flux, but has a lower contrast with the underlying dust grain continuum due to its wider bandwidth.
The \ce{OH} (0-0) band at $\lambda\simeq 3080$\AA{} generally has higher flux, but unless observed from space the high terrestrial atmospheric absorption severely diminishes the observed flux.
Strong \ce{CO+} emission bands have also been observed in a number of comets \citep[the strongest ones are the (3-0) and (2-0) $\mathrm{A^2\Pi-X^2\Sigma}$; ][]{Biver2018}.
However this is normally significantly weaker than the observed \ce{CN} emission, with detection in only less than 20 comets.
Given the ubiquitous signature of \ce{CN} emission in Solar System comets, we have embarked on an archival search for this species in the exocomets around \bp{}.
Detection of this species would provide a solid link between cometary bodies in our Solar System and the \bp{} system.

We detail the observations in Sect.~\ref{sec:data}, along with the correction for the echelle blaze function and normalization of the spectra to a continuum.
Our search for \ce{CN} relies on a cross correlation search with a theoretical absorption spectrum for the molecule in Sect.~\ref{sect:CNsearch}, and the discussion and interpretation is in Sect.~\ref{sec:discuss} with our conclusions in Sect.~\ref{sec:conclusion}.

\section{Data description and analysis}\label{sec:data}

For the study in this paper, we use all the publicly available data from the \ac{harps} and the ESO 3.6m telescope in Chile.

\subsection{HARPS observations}

We collated all publicly available HARPS observations of \bp{} obtained between October 2003 and March 2020.
Although these observations cover a large baseline, the time between visits varies between 1 day and approximately 2.2 years.
Similarly, the number of observations per epoch spans a wide range, between one and 356 exposures, while the exposure times vary between 12 and 900 seconds.
This results in an average SNR across order 0 - 10 (order numbers for HARPS will follow the labelling in the fits header -- corresponding to 3780\AA{} and 4075\AA{}) per exposure between 1 and 430, and a nightly SNR between 63 and 2000.
To avoid potential edge effects from order merging,  we used the E2DS files downloaded directly from the ESO archive for our analysis.

\subsection{Initial data processing}
The analysis was started from the 2d extracted spectra (E2DS files).
Each order was treated separately, with the main focus on orders 3 and 7 as these orders contain the main \ce{CN} bandhead as well as the \ion{Ca}{ii} H line, which is used to identify the exocomets.
All the spectra were transformed into the stellar rest-frame and onto a common wavelength grid using linear interpolation.

After all the spectra were all transformed into the stellar rest frame, an average nightly spectrum was made from all spectra within a night for each of the nights of observations. This was done after correcting for variations in the effective blaze profile (for example due to instrumental changes, differential slit or fibre losses, atmospheric absorption, etc.), as outlined below. Note that after this correction, the data still have the overall instrumental blaze-profile included.

\subsection{Effective Blaze function correction}\label{sect:blaze}

The night of 06 May 2018, which has a high signal-to-noise ratio, was chosen to act as an initial estimate for the blaze function.
The individual spectra from this night were averaged together to further boost the signal-to-noise ratio.
Subsequently, each individual spectrum was divided by this reference spectrum to highlight the changes in the effective profile.
This ratio spectrum was subsequently binned using a bin-width of 81 pixels using a robust algorithm to remove outliers within each bin, clipping points more than 2.7 Median Absolute Deviations (MAD) away from the median of each bin.
This binned ratio was then fitted using a second order polynomial.
The original spectrum was finally divided by this polynomial to correct for the changes.
Note that the change of fibre for HARPS in June 2015 (JD$\sim$2,457,180) resulted in an overall change in the blaze profile, and therefore, the spectra before and after this date were separately post-processed as outlined below in Sect.~\ref{sect:starcor}.

\subsection{Combining the spectra and flattening them}\label{sect:comb}

After the blaze-variation corrections, all the spectra within a night are combined in order to further boost the signal-to-noise and improve the detection limits.
Each individual spectrum was weighted using the average signal-to-noise ratio provided in the header for orders 0 and 11 (3780\AA{} - 4075\AA{}).

\subsection{Removing the stellar spectrum}\label{sect:starcor}

The processing up to this stage removes most of the variations in the effective blaze function, but the overall stellar spectrum and shape of the blaze profile, are still present and need to be removed. 
We first perform a second iteration of blaze corrections using a stellar spectrum constructed by averaging the spectra from periods with a low comet activity (see Sect.~\ref{sect:FEBid}), similar to the method before, however, now using a wider binsize of 150 pixels and clipping outliers at 4.5 MAD away from the median.
This is done separately pre- and post-fibre change. The correction is done similar to Sect.~\ref{sect:blaze}, however, an 8th order polynomial was used.

After this post-processing the overall (residual) blaze profile and stellar spectrum  needs to be removed.
Care must be taken to prevent the removal of (potential) narrow features from the circumstellar disk (as seen in, for example, the \ce{Fe} and \ion{Ca}{II} lines).
This is done by combining the nightly spectra without significant cometary activity (see Sect.~\ref{sect:FEBid}) and binning this stacked spectrum in wavelength in bins of 21 pixels.
Before binning, the spectrum in each bin was median absolute deviation (MAD) clipped to within $\pm$4.5 MAD of the median of the flux in the bin.
The clipped average of the fluxes within each bin was used.
Subsequently, the overall spectrum of the star and (residual) blaze was determined by cubic interpolation across the binned points, and divided out of the spectrum for each night.
As before, this was done separately pre- and post-fibre change.

\section{Searches for \texorpdfstring{\ce{CN}}{CN}}\label{sect:CNsearch}

\begin{figure}
    \begin{centering}
        \includegraphics[width=\columnwidth]{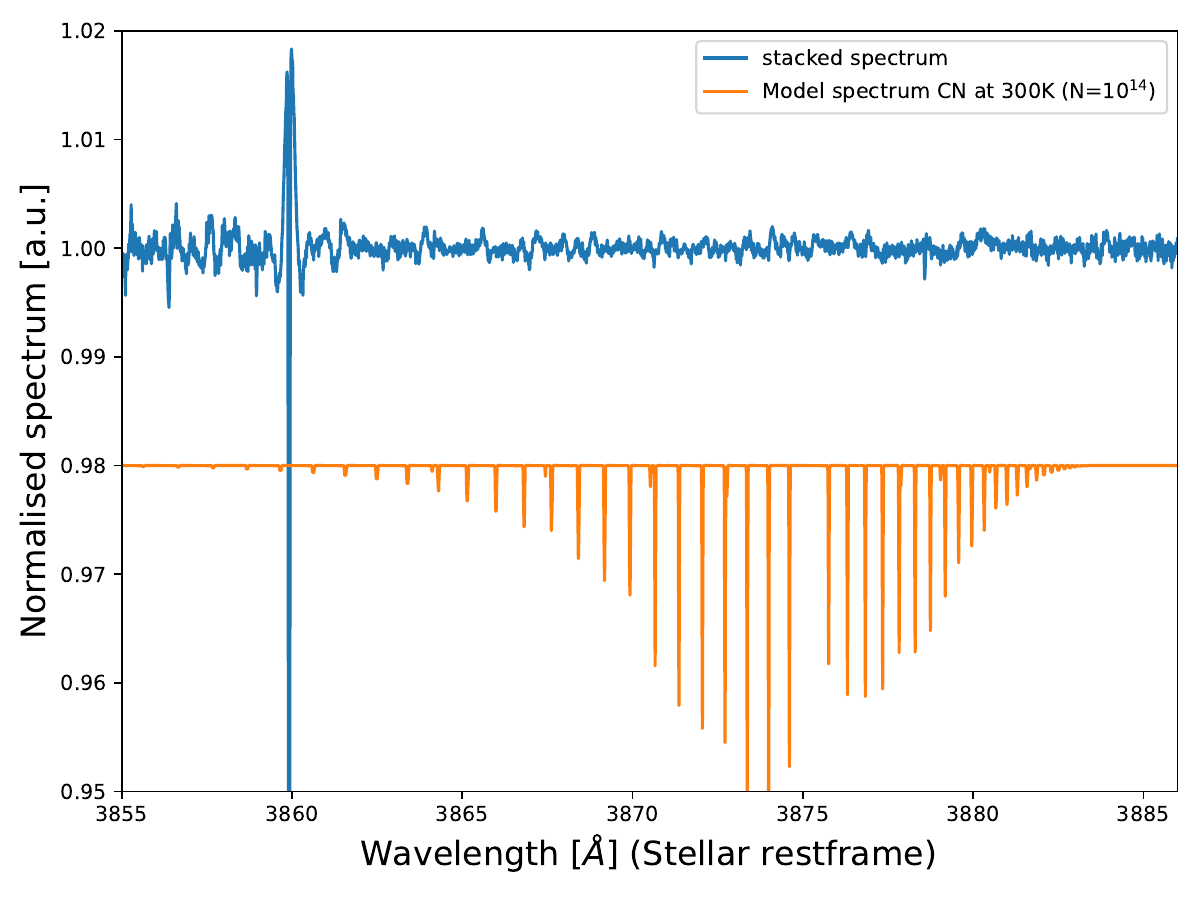}
        \caption{Plot of the stacked spectrum in the stellar restframe (blue) with a CN model for $T=$300~K and N=10$^{14}$\,cm$^{-2}$ overplotted in orange.
        The large signal around 3860\AA{} is due to the residual from a circumstellar \ion{Fe}{I} line.}
        \label{fig:spec_CN_stellar_frame}
        \script{plot_spec_mean.py}
    \end{centering}
\end{figure}

After the initial data analysis, we proceed to with our search for lines from \ce{CN}.
Although we are interested in lines from the exocomets, we first co-add the data from all the individual epochs, in the stellar rest-frame, to obtain a deep search for circumstellar \ce{CN} absorption.
These coadded spectra are shown in Fig.~\ref{fig:spec_CN_stellar_frame}.
No obvious feature from the \ce{CN} band jumps out, and we need to combine the signal from the lines within the \ce{CN} band to improve the detection limits.
Since the exocomets have a clear radial velocity with respect to \bp{}, and since each exocomet will have its own distinct velocity, we selected the 36 epochs with the strongest exocometary features, a seen in the \ion{Ca}{ii} H line. 

\subsection{Identifying exocomets from \texorpdfstring{\ion{Ca}{ii}}{CaII} H line}\label{sect:FEBid}

\begin{figure}
    \begin{centering}
        \includegraphics[width=\columnwidth]{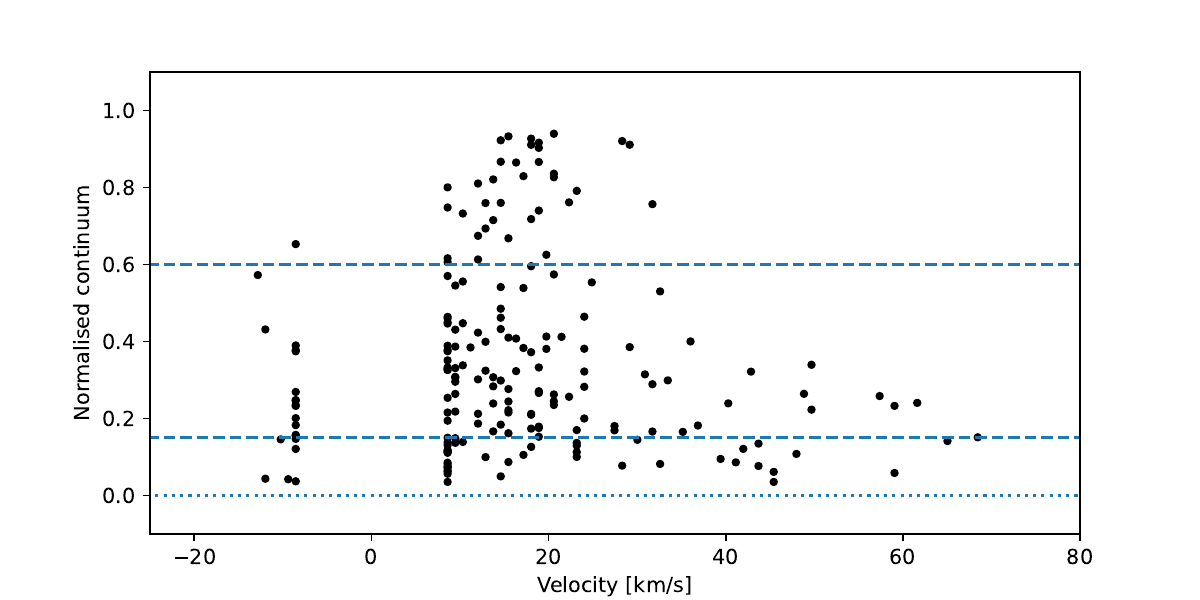}
        \caption{Peaks and radial velocities of exocomets identified in the stellar spectrum.
        The peaks are normalised such that 1 represents total absorption in the spectrum.
        Peaks closer than 10 \kms{} are masked out due to contamination from the circumstellar material.
        Peak values above 0.6 are counted as exocomets and those below 0.15 as stellar only.}
        \label{fig:FEB_velocity}
        \script{plot_FEB_velocity_depth.py}
    \end{centering}
\end{figure}

The comet features in the \ion{Ca}{ii} H-line ($\lambda=3968$ \AA) were identified after removing the overall stellar line-profile and avoiding the absorption from the circumstellar disk by excluding the central 8 \kms{}.
The deepest absorption feature in each observing night was selected, and the location and depth stored.
For the comet sample we selected features deeper than 60\% while spectra with features below 15\% were flagged as `low activity'.
The distribution of feature locations and depths is shown in Fig.~\ref{fig:FEB_velocity}. The dates, velocities and depths of the selected exocomets can be found in Tab.~\ref{tab:FEB_data}

\begin{figure}
    \begin{centering}
        \includegraphics[width=\columnwidth]{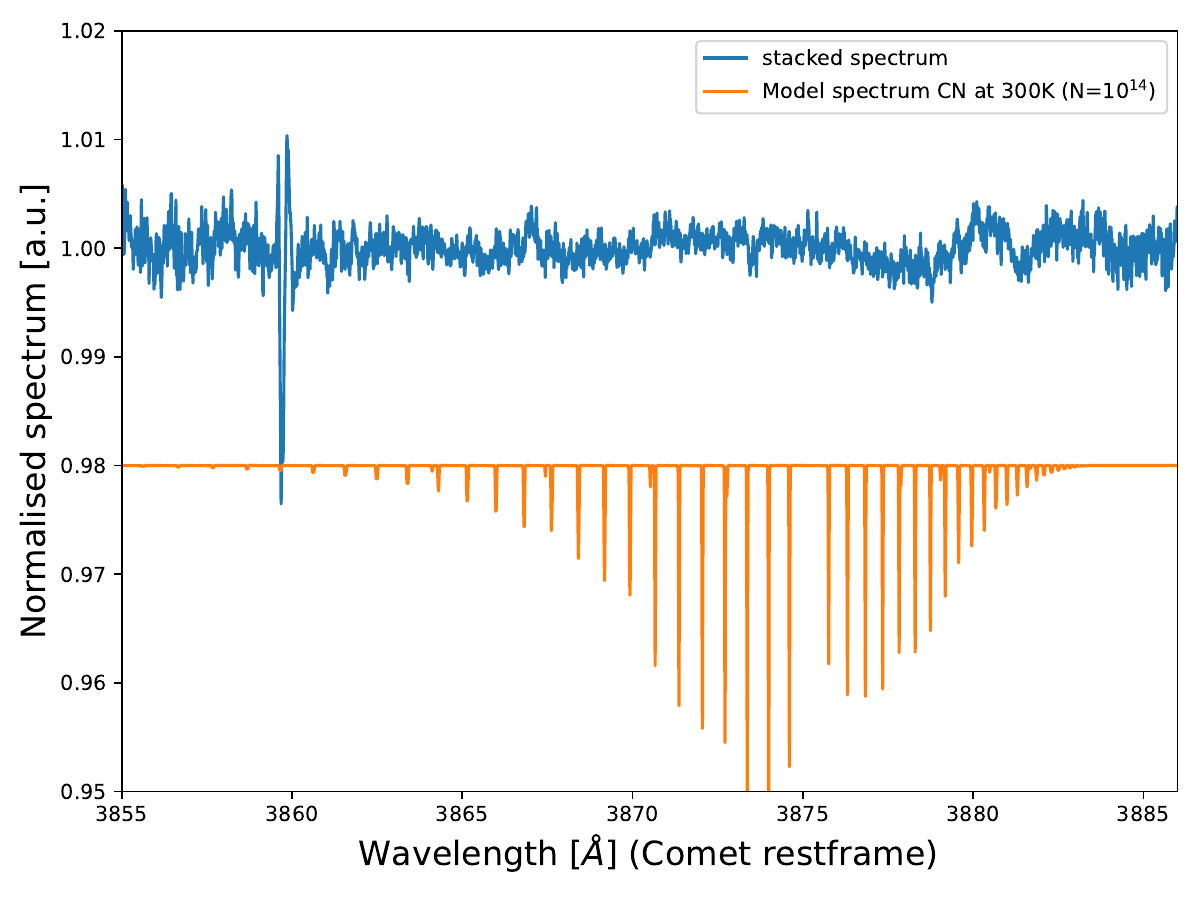}
        \caption{Similar to Fig.~\ref{fig:spec_CN_stellar_frame} but spectra are stacked in the rest frame of the comet. The large signal around 3860\AA{} is due to the residual from a circumstellar \ion{Fe}{I} line.}
        \label{fig:spec_CN_comet_frame}
        \script{plot_spec_mean.py}
    \end{centering}
\end{figure}

The spectra for the strong comet sample were coadded, and the combined spectrum for the HARPS data is shown in Fig.~\ref{fig:spec_CN_comet_frame}.
As with the circumstellar \ce{CN}, there are no obvious lines from \ce{CN} visible, and we will need to combine multiple lines to improve the signal-to-noise ratio, and see if there is any \ce{CN} present. 

\subsection{Cross-correlation with \texorpdfstring{\ce{CN}}{CN} models}\label{sect:CCF}

\begin{figure}
    \begin{centering}
        \includegraphics[width=\columnwidth]{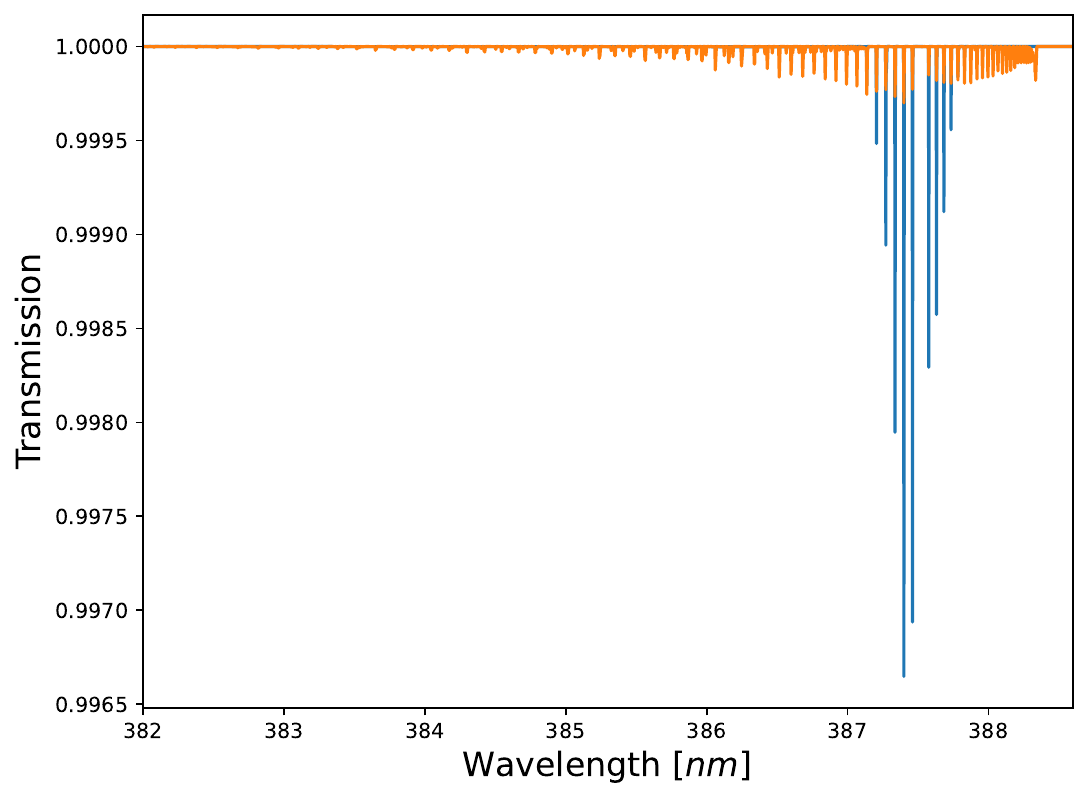}
        \caption{Model absorption spectrum for \ce{CN} at 30\,K (blue line) and 2000\,K (orange line). The  column densities are $N=10^{12}$\,cm$^{-2}$ and $N=10^{13}$\,cm$^{-2}$, respectively. }
        \label{fig:CN_theory}
        \script{plot_two_CN_temps.py}
    \end{centering}
\end{figure}

\begin{figure}
    \begin{centering}
        \includegraphics[width=\columnwidth]{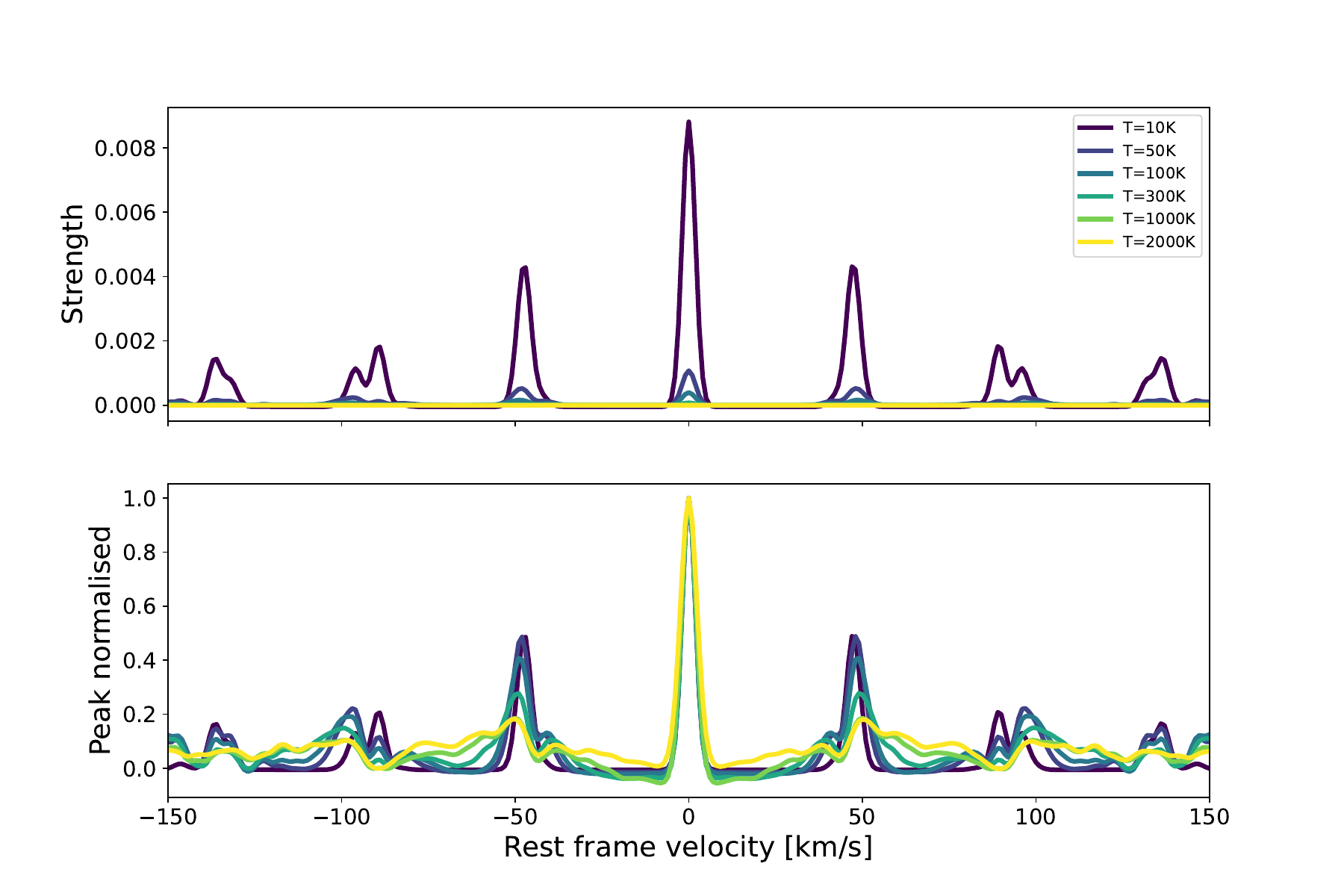}
        \caption{\ce{CN} autocorrelation function for different temperatures. The aliasing at $\sim50$\,\kms{} is clearly visible}
        \label{fig:CN_ACF}
        \script{plot_CN_ACF.py}
    \end{centering}
\end{figure}

To combine the signal from the many \ce{CN} lines, we cross-correlate the data for each night with a grid of model of \ce{CN}.
Cross-sections for \ce{CN} were obtained from ExoMol\footnote{\url{https://www.exomol.com}} \citep{Tennyson12,Tennyson16,Brooke14} with molecular energy levels assumed to be in thermal equilibrium, i.e.\ Boltzmann distributed.
We made models for a wide range of temperatures (T$=$\{10, 20, 30, 50, 100, 200, 300, 500, 1000, 2000, 3000\} K) and column densities\footnote{Note that technically, for optically thin lines, this assumes a filling factor $f=1$.
Since $f$ is unknown, we actually measure N$f$} (N=\{$10^{11} \ldots 10^{15}$\} cm$^{-2}$.
Each of the models was then convolved to include an effective velocity dispersion using a Gaussian for FWHM=\{2.9, 5, 10, 15, 20\} \kms{}.
A value of 2.9 \kms{} corresponds approximately to the instrumental resolution of HARPS.
The theoretical spectra of \ce{CN} for several different temperatures are shown in Fig.~\ref{fig:CN_theory}.

\begin{figure}
    \begin{centering}
        \includegraphics[width=\columnwidth]{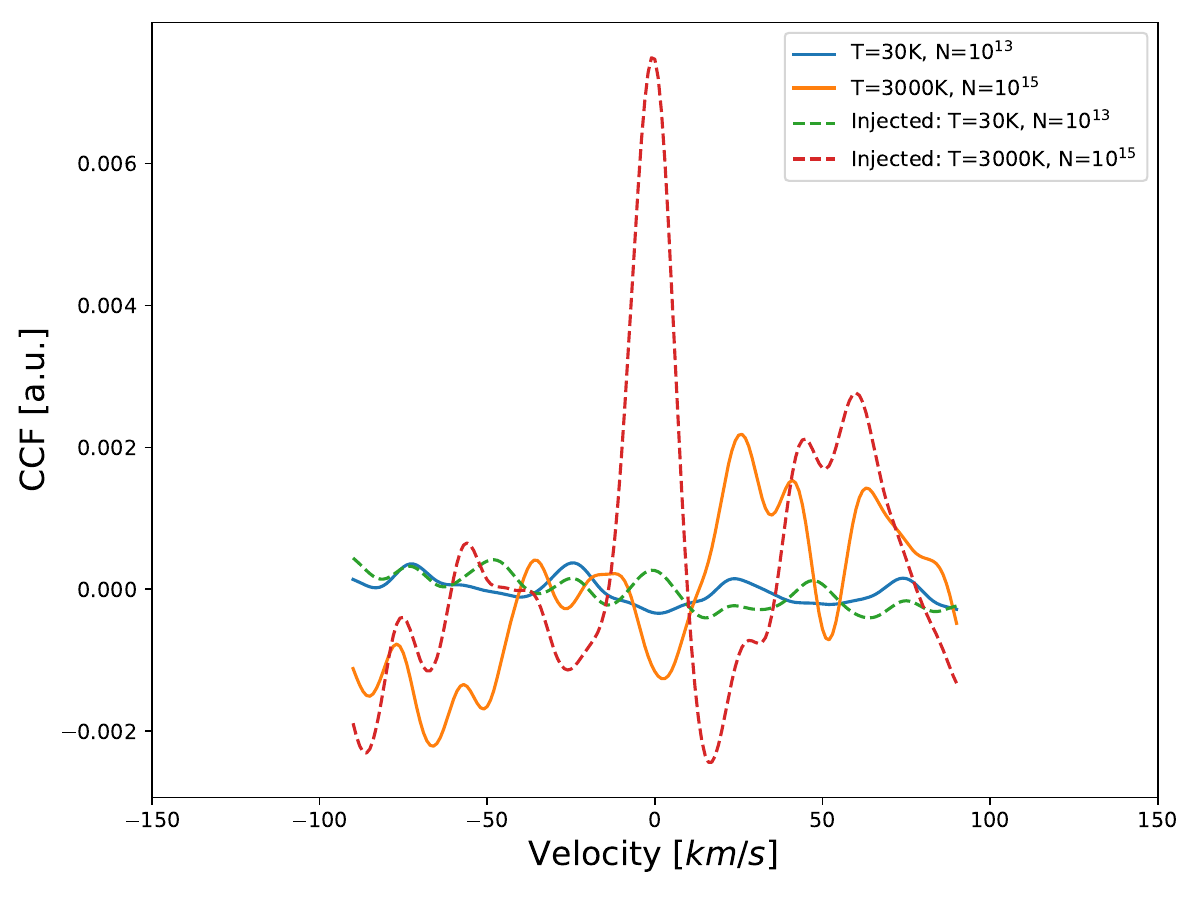}
        \caption{Cross correlation for different number densities of \ce{CN} averaged over many spectra.}
        \label{fig:ccf_mean}
        \script{plot_ccf_mean.py}
    \end{centering}
\end{figure}

When creating the CCFs, the region around the narrow, circumstellar \ion{Fe}{I} lines in this wavelength region (see Figs.~\ref{fig:spec_CN_stellar_frame}~and~\ref{fig:spec_CN_comet_frame}) were masked to avoid contamination, especially for the higher temperature CN models.
We masked the lines detected in \citep{Kiefer2018}, using a width of $\pm0.35$\AA{} for the \ion{Fe}{I} 3859.9\AA{} line, while for the other lines a width of $\pm0.15$\AA{} was used.
Even after cross-correlating, the stacked CCFs (see Fig.~\ref{fig:ccf_mean}) do not show any clear signals, and we therefore proceed to estimate our detection limits.

\begin{figure}
    \begin{centering}
        \includegraphics[width=\columnwidth]{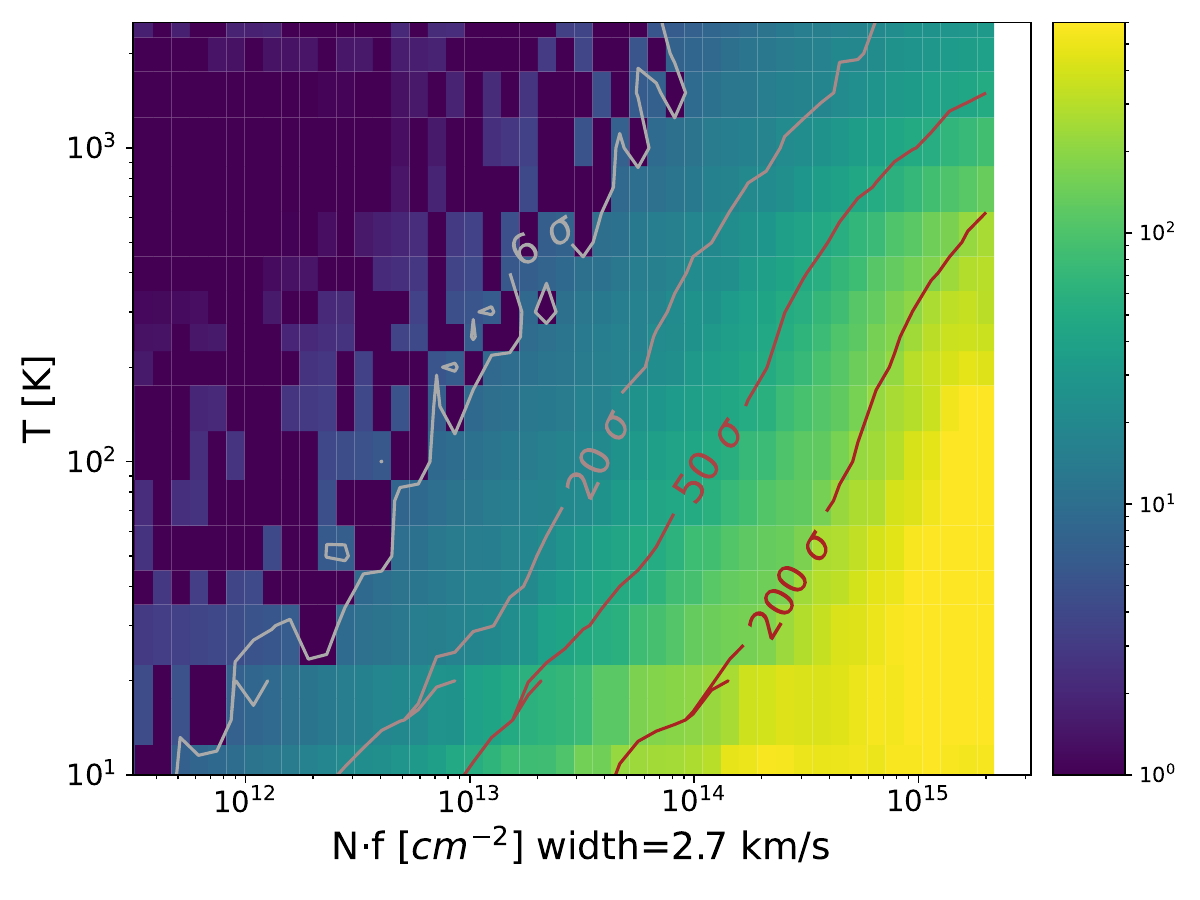}
        \caption{\ce{CN} measurement limits in the stellar rest frame for a \ce{CN} line width of 2.7 \kms{}. Contours show signal-to-noise upper limits.}
        \label{fig:CN_stellar_frame}
        \script{plot_HARPS_CN_stellar_frame.py}
    \end{centering}
\end{figure}

\begin{figure*}
    \begin{centering}
        \includegraphics[width=\linewidth]{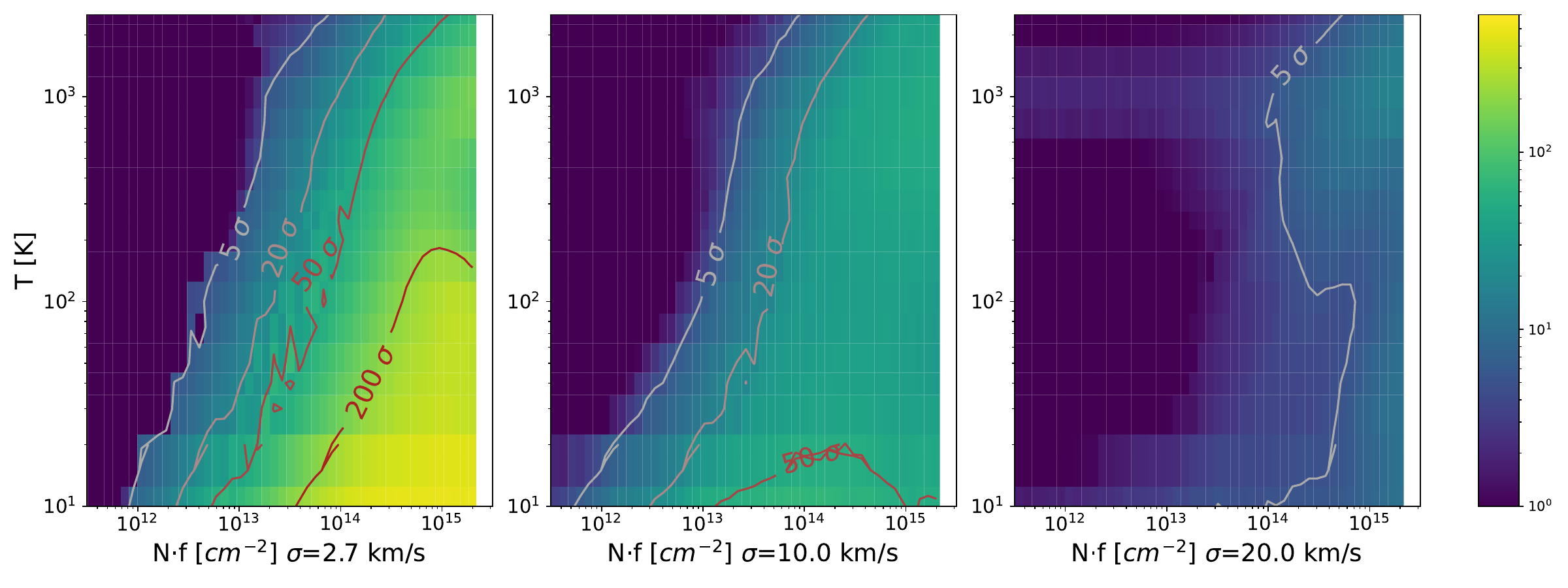}
        \caption{\ce{CN} measurement limits in the exocomet frame for a \ce{CN} line width of 2.7, 10 and 20 \kms{}. Contours show signal-to-noise upper limits.}
        \label{fig:CN_exocomet_frame}
        \script{plot_HARPS_CN_exocomet_frame.py}
    \end{centering}
\end{figure*}

\subsection{Estimating the \texorpdfstring{\ce{CN}}{CN} column density limits}\label{sect:CNlim}

To estimate the limits on the \ce{CN} detection, we perform injection and recovery tests.
We injected models for different temperatures, column densities and velocity broadening into the nightly stacked spectra (Sect.~\ref{sect:comb}), and then processed them in the same way as the original data (i.e., correct for residual blaze variations, removal of the stellar envelope and cross-correlation).
This was done for all spectra with the injected signal in the stellar rest frame as well as for the sample of spectra with strong cometary activity in the rest frame of the strongest exocometary feature per night.

The individual CCFs show some overall slopes and offsets, and we fit each individual CCF with a combination of a second-order polynomial baseline and a Gaussian with the position fixed to 0 \kms{} in its rest frame.
The fit has five free parameters: the amplitude of the peak, the width of the peak, and the offset, slope, and intercept of the baseline.
To avoid the aliasing peaks of the \ce{CN} band (see Fig.~\ref{fig:CN_ACF}), we only fit the central $\pm$35 \kms{} of the CCF.
Fitting is done using the {\tt curve\_fit} function from SciPy \citep{2020SciPy-NMeth}.
We correct each of the individual CCFs for the baseline and then stack the corrected CCFs to obtain our master CCF that we use for determining the upper limit.
The noise on each point in the master CCF is determined by taking the standard deviation in time for each pixel and dividing that by the square root of the number of points.

This master CCF is then fitted in the same way as the nightly CCFs, and the uncertainties on the parameters is taken directly from {\tt curve\_fit} assuming no covariance between the parameters.
The final detection significance for the injected models is determined by the amplitude divided by its uncertainty, and shown in Fig.~\ref{fig:CN_stellar_frame} for the circumstellar gas and in Fig.~\ref{fig:CN_exocomet_frame} for the strongest exocomet features.

\section{Discussion}\label{sec:discuss}

As discussed above in the introduction, most studies of Solar System comets observe the \ce{CN} molecules via optical fluorescence. 
From measurements and simple physical arguments the \ce{CN} absorption or emission is optically thin everywhere.
This means a flux from part of the coma can immediately be converted to a column density using a photon scattering efficiency for the molecular band.

\ce{CN} in Solar System comets comes from two sources.
Sublimation of nucleus near-surface ices releases \ce{HCN}, which then photo-dissociates into \ce{CN}.
This in turn eventually photo-dissociates into the component atoms:

$$\ce{HCN} +\gamma \rightarrow \ce{H}+\ce{CN}$$
$$\ce{CN} + \gamma \rightarrow \ce{C}+\ce{N}$$

Since observations of Comet Halley \citep{AHearn1986}, it has also been known that dust grains ejected from the nucleus also directly release \ce{CN} into the coma.
However, \ce{CN} production in comets is usually dominated by nuclear ice sublimation \citep{Fray2005}.
The resulting \ce{CN} coma will have a true space density and apparent column density distribution centrally peaked at the comet nucleus in this case.

For Solar System comets, measured \ce{CN} column densities are often converted to an overall production rate by the comet of $Q_{\ce{CN}}$ molecules per second via an analytical model known as the ``Haser'' model \citep{Haser2020}.
This model gives the on-sky column density as a function of projected radial distance form the nucleus using only three parameters; the effective destruction scale length $l_p$ of the parents of \ce{CN} (\ce{HCN}+dust), the photodissociation scale length $l_{\ce{CN}}$ of \ce{CN}, and the gas outflow velocity $v$.
The Haser model is somewhat simplistic but it represents the data well for Solar System comets observed at typical distances from the Sun (1--3 au).

\subsection{Example comet - 1P/Halley}

We take the observed activity of 1P/Halley in 1985/1986 as a reference point for our estimates.
Although most Solar System comets are observed to be are far less active than Halley, some Long-Period comets show activity in excess of Halley.
Given that we will be dominated by the most active comets at \bp{}, this seems reasonable as something that can be later scaled to the most active comets such as Hale-Bopp \citep{Schleicher2024}.
When Halley was approximately one au from the Sun, it was measured to have $Q_{\ce{CN}}\simeq 10^{27}$ molecules per second using Haser models \citep{AHearn1995}.
The assumed outflow velocity in the models was generally  $0.85 - 1$ \kms{} at this distance.
For Solar System comets, the normal Haser scale lengths used are \citep{Cochran86,AHearn1995}:

$$lp=1.7\times10^4r_h^2\:{\rm km}$$
$$l_{\ce{CN}}=2-3\times10^5r_h^2\:{\rm km}$$

Both are assumed to scale with heliocentric distance $r_h^2$ as photodissociation or thermal desorption is assumed the main physical pathway for both the parent and \ce{CN}.
The lifetime of \ce{HCN} for photodissociation into \ce{H}+\ce{CN} at 1~au has been calculated to be $(3.2-7.7)\times10^4$ sec, with lifetime of \ce{CN} dissociating into \ce{C}+\ce{N} being $(1.3-3.1)\times10^5$ sec \citep{Huebner92}, depending on solar activity levels.
These are consistent with the measured scale lengths above if the parents of \ce{CN} are ejected at approximately $0.5$ \kms{}, and we take into account the excess photodissociation energy giving \ce{CN} an extra kick of $\simeq0.9$ \kms{} \citep{Fray2005}. 

\subsection{1P/Halley at \texorpdfstring{\bp{}}{Beta Pictoris}}

Photodissociation is  driven by the UV photon flux at the comet, which in turn is dominated by the Ly-$\alpha$ line flux \citep{Combi80}.
The  Ly-$\alpha$ emission from \bp{} has been measured by \citet{Wilson17} via HST+COS.
Using their ISM and CSM corrected Voigt line profile parameters, we integrated over $\pm600$ \kms{} and corrected for the distance of \bp{} to derive an intrinsic Ly-$\alpha$ luminosity of $1.4\times10^{30}$ ergs sec$^{-1}$.

The solar Ly-$\alpha$ flux varies by $\sim 50\% $ over the Solar cycle, with a mean value at Earth of $4.7\times10^{11}$ phot s$^{-1}$ cm$^{-2}$ \citep{Woods00}.
The resulting mean solar Ly-$\alpha$ luminosity is $2.1\times10^{28}$ ergs sec$^{-1}$, a factor of $\sim 67$ lower than \bp{}.
Hence the photodissociation lifetime of \ce{CN} at 1 au from \bp{} of an `exoHalley' is approximately 67 times shorter, for example around $3000$ sec.

In steady-state sublimation, the sublimation rate is proportional to the photon flux at the comet.
For \bp{} $L=8.7L_\odot$, so assuming all \ce{CN} production scales with stellar luminosity, an exoHalley has a larger \ce{CN} production rate of $Q_{\ce{CN}}\sim 9\times10^{27}$ molecules per second.
The total number of \ce{CN} molecules in the coma at 1~au at any given time would therefore be:

$$N_{\ce{CN}}=Q_{\ce{CN}} \tau_{\ce{CN}}$$

$$N_{\ce{CN}}\sim 9\times10^{27} \cdot  3000 \sim 3\times10^{31}\:{\rm molecules}$$

The effective cross section of the coma will depend on how much is projected against the disk of \bp{}.
The radius of the \ce{CN} coma is then given by
$$r\sim v_{\ce{CN}} \tau_{\ce{CN}}$$
The outflow velocity of the parent \ce{HCN} will depend on the nuclear surface temperature, which will increase as a comet moves towards its parent star.
\citet{Biver2002} found a nuclear gas outflow velocity $v\simeq1.0  r_h^{-0.4}$ \kms{} from sub-mm measurements of Comet Hale-Bopp, close to the canonical relationship of $r_h^{-0.5}$.
\citet{Harris2002} notes lower activity similar to Halley results in smaller outflow velocities.
The excess photodissociation velocity of \ce{CN} will not change.
Therefore gas outflow speeds are expected to be $\sim 1-2$ \kms{} at 1 AU, and the \ce{CN} coma radius will be:
$$r\sim (1-2) \times 3000 \leq 6\times 10^{3}\:{\rm km}.$$
This is much smaller than a stellar radius, so we can assume for an exoHalley that all of the coma is projected into the stellar disk as seen from Earth.

\bp{} has $R=1.8R_\odot\equiv 1.3\times 10^{11}$ cm.
Given that the coma is optically thin, and we do not resolve the star, the effective average column density towards \bp{} is: 

$$n\sim\frac{N_{\ce{CN}}}{\pi R_*^2}$$

$$n\sim\frac{3\times10^{31}}{\pi\cdot 1.7\times10^{22}}\sim 6\times10^{8}\:{\rm cm^{-2}},$$
therefore an exoHalley at 1~au from \bp{} would result in a \ce{CN} column density of $10^{8}-10^{9}$ cm$^{-2}$.
This is $\geq100$ times smaller than our measured upper limits above.

\subsection{Raising the \texorpdfstring{\ce{CN}}{CN} production rates}

There are several ways exocomets at \bp{} may exhibit larger \ce{CN} production rates.
A larger comet gives more surface area for sublimation.
 Comet Halley's nucleus has an effective radius of $5.5$ km \citep{Keller1987}, whereas Comet Hale-Bopp was one of the most intrinsically active comets on record with an extremely large nucleus of radius around $14-21$ km \citep{Weaver1997}.
However most of the apparent brightness was due to dust emission, and its only had $Q_{\ce{CN}}\simeq 3\times10^{27}$ molecules per second \citep{Schleicher2024}.
In this case, we might anticipate the most active comets in the Solar System to have $Q_{\ce{CN}}\sim 10^{28}$ molecules per second, implying a commensurate increase of a factor 10 in column density $n$ over an exoHalley at \bp{}.

Most comets only have a fraction of their surface area undergoing mass-loss, due to the effect of a thermally insulating dust crust.
Halley itself only has $\sim20$\% of the sunlit surface active \citep{Keller1987}.
Yet some comets are known as hyperactive, as the measured gas productions rates would require $\geq 100$\% of the nucleus surface to be active \citep{Sunshine2021}.
This is believed to be due to the release of ice-rich dust grains from the nucleus, which sublimate and act as a secondary source of gas.
Assuming this is possible for exocomets at \bp{} would increase $N_{\ce{CN}}$ and $n$ by a factor of around five.

The closer a comet to its parent star, the greater the heating and corresponding production rate.
When insolation completely controls the nuclear surface energy balance we expect
$Q_{\ce{CN}}\propto r_h^{-2}$.
Studies of Halley and Hale-Bopp showed power laws close to this \citep[e.g. ][]{Biver2002}.
But the photodissociation lifetime $\tau_{\ce{CN}}\propto r_h^{2}$, so theoretically:
$$N_{\ce{CN}}=\frac{Q_{\ce{CN}}(1~\rm{au})}{r_h^2}\cdot  \tau_{\ce{CN}} (1~\rm{au}) r_h^2 =\:{\rm constant},$$
hence the total amount of \ce{CN} in the coma should be approximately independent of stellar distance, and the total cross-section would not change.

Finally, comets have impulsive outbursts, where the amount of material ejected into the coma increases significantly over a few hours or less.
The general causes are unclear, but there is evidence for several physical mechanisms that cause them \citep{Kelley2021,Muller2024}.
``Normal'' outbursts generally result in a brightness increases of 1--5 magnitudes, with smaller outbursts being more common.
As the apparent brightness is directly proportional to the amount of material (gas+dust) in the coma, this implies short term increases in $N_{\ce{CN}}$ and $n$ of factors ranging from 2 to 100.
Large outbursts are often due to partial or total disruption of a comet and can occur when a comet is closer to the Sun.
Hence there could be a correlation with faster moving \bp{} exocomets nearer their periastra producing greater measurable column densities of \ce{CN}.

In summary, mechanisms exist by which the \ce{CN} column density of an exocomet at \bp{} could exceed $10^{9}$ cm$^{-2}$.
A larger or more hyperactive nucleus could increase $n$ by a factor of ten.
Even larger increases are possible with an outburst or comet disruption, but the upper boundary is likely to be near $n\sim10^{11}$ cm$^{-2}$.

\section{Conclusions}\label{sec:conclusion}

We have combined over 1000 spectra of \bp{} over a 20 year period, identified two groups, one with deep exocomet signals, and another to act as a stellar reference.
We produce a normalised spectrum towards \bp{} with the stellar spectral features removed, and then aligned the spectra to the exocomet rest frames and perform a search for \ce{CN} using cross correlation.

We do not detect \ce{CN} both in the stellar rest frame and in the exocomet rest frame.
Injection and recovery tests show that for the circumstellar \ce{CN} we are sensitive to $10^{12.5}$ cm$^{-2}$ below 80~K and below $10^{14}$cm$^{-2}$ up to 1800~K.
In the exocomet rest frames, sensitivity is somewhat similar.
Below 100~K the sensitivity is $10^{12}$cm$^{-2}$ and above from 100 - 1800~K, is $10^{13}$cm$^{-2}$.
This is a factor of 10 to 100 too high to detect a single Hale-Bopp equivalent comet with a \ce{CN} coma.
The assumptions are for steady state comets, so individual outbursts could build this number up, and multiple transiting exocomets can increase the absorption cross section.

\bp{} is the brightest star with known exocomets seen both in transit and spectroscopy, and represents the best candidate for exocomet characterisation.
A more dedicated observing campaign over several years using dedicated spectrographs can reach the equivalent Solar System limits for \ce{CN}.
PLATO will carry out a dedicated long stare campaign that will observe the star with sub-millimagnitude photometry over two to three years.
This presents an excellent opportunity for coordinated spectroscopic campaigns where the exocomet population of \bp{} can be characterised, with \ce{CN} and ideally other molecules known from Solar System comets to be identified in another stellar system.

\section*{Data availability}

An online repository with materials used in this work is available at \url{https://github.com/mkenworthy/BetaPicCN} using the {\tt showyourwork!} package \citep{Luger2021}.

\begin{acknowledgements}

The authors thank the Lorentz Centre at Leiden University for organising the workshop ``Exocomets: Understanding the Composition of Planetary Building Blocks'' which led to the study in this paper. 
Part of this work was supported by the German \emph{Deut\-sche For\-schungs\-ge\-mein\-schaft, DFG\/} project number Ts~17/2--1.
AF acknowledges support from STFC award ST/T00021X/1.
EdM acknowledges support from STFC award ST/X00094X/1.
We made use of the {\tt Python} programming language \citep{rossum1995} and the open-source {\tt Python} packages {\tt numpy} \citep{walt2011}, {\tt matplotlib} \citep{hunter2007}, {\tt astropy} \citep{astropy2013}.

\end{acknowledgements}

\bibliographystyle{aa}
\bibliography{bib.bib}

\begin{thebibliography}{68}
\expandafter\ifx\csname natexlab\endcsname\relax\def\natexlab#1{#1}\fi

\bibitem[{{A'Hearn} {et~al.}(1986){A'Hearn}, {Hoban}, {Birch}, {Bowers},
  {Martin}, \& {Klinglesmith}}]{AHearn1986}
{A'Hearn}, M.~F., {Hoban}, S., {Birch}, P.~V., {et~al.} 1986, \nat, 324, 649

\bibitem[{{A'Hearn} {et~al.}(1995){A'Hearn}, {Millis}, {Schleicher}, {Osip}, \&
  {Birch}}]{AHearn1995}
{A'Hearn}, M.~F., {Millis}, R.~C., {Schleicher}, D.~O., {Osip}, D.~J., \&
  {Birch}, P.~V. 1995, \icarus, 118, 223

\bibitem[{{Astropy Collaboration} {et~al.}(2013){Astropy Collaboration},
  {Robitaille}, {Tollerud}, {Greenfield}, {Droettboom}, {Bray}, {Aldcroft},
  {Davis}, {Ginsburg}, {Price-Whelan}, {Kerzendorf}, {Conley}, {Crighton},
  {Barbary}, {Muna}, {Ferguson}, {Grollier}, {Parikh}, {Nair}, {Unther},
  {Deil}, {Woillez}, {Conseil}, {Kramer}, {Turner}, {Singer}, {Fox}, {Weaver},
  {Zabalza}, {Edwards}, {Azalee Bostroem}, {Burke}, {Casey}, {Crawford},
  {Dencheva}, {Ely}, {Jenness}, {Labrie}, {Lim}, {Pierfederici}, {Pontzen},
  {Ptak}, {Refsdal}, {Servillat}, \& {Streicher}}]{astropy2013}
{Astropy Collaboration}, {Robitaille}, T.~P., {Tollerud}, E.~J., {et~al.} 2013,
  \aap, 558, A33

\bibitem[{{Bendahan-West} {et~al.}(2024){Bendahan-West}, {Kennedy}, {Brown}, \&
  {Str{\o}m}}]{BendahanWest24}
{Bendahan-West}, R., {Kennedy}, G.~M., {Brown}, D. J.~A., \& {Str{\o}m}, P.~A.
  2024, \mnras [\eprint[arXiv]{2412.13253}]

\bibitem[{{Beust} {et~al.}(1996){Beust}, {Lagrange}, {Plazy}, \&
  {Mouillet}}]{Beust96}
{Beust}, H., {Lagrange}, A.~M., {Plazy}, F., \& {Mouillet}, D. 1996, \aap, 310,
  181

\bibitem[{{Beust} {et~al.}(1991){Beust}, {Vidal-Madjar}, {Lagrange-Henri}, \&
  {Ferlet}}]{Beust91}
{Beust}, H., {Vidal-Madjar}, A., {Lagrange-Henri}, A.~M., \& {Ferlet}, R. 1991,
  \aap, 241, 488

\bibitem[{{Biver} {et~al.}(2002){Biver}, {Bockel{\'e}e-Morvan}, {Colom},
  {Crovisier}, {Henry}, {Lellouch}, {Winnberg}, {Johansson}, {Gunnarsson},
  {Rickman}, {Rantakyr{\"o}}, {Davies}, {Dent}, {Paubert}, {Moreno}, {Wink},
  {Despois}, {Benford}, {Gardner}, {Lis}, {Mehringer}, {Phillips}, \&
  {Rauer}}]{Biver2002}
{Biver}, N., {Bockel{\'e}e-Morvan}, D., {Colom}, P., {et~al.} 2002, Earth Moon
  and Planets, 90, 5

\bibitem[{{Biver} {et~al.}(2018){Biver}, {Bockel{\'e}e-Morvan}, {Paubert},
  {Moreno}, {Crovisier}, {Boissier}, {Bertrand}, {Boussier}, {Kugel}, {McKay},
  {Dello Russo}, \& {DiSanti}}]{Biver2018}
{Biver}, N., {Bockel{\'e}e-Morvan}, D., {Paubert}, G., {et~al.} 2018, \aap,
  619, A127

\bibitem[{Borucki {et~al.}(2010)Borucki, Koch, Basri, Batalha, Brown, Caldwell,
  Caldwell, Christensen-Dalsgaard, Cochran, DeVore, Dunham, Dupree, Gautier,
  Geary, Gilliland, Gould, Howell, Jenkins, Kondo, Latham, Marcy, Meibom,
  Kjeldsen, Lissauer, Monet, Morrison, Sasselov, Tarter, Boss, Brownlee, Owen,
  Buzasi, Charbonneau, Doyle, Fortney, Ford, Holman, Seager, Steffen, Welsh,
  Rowe, Anderson, Buchhave, Ciardi, Walkowicz, Sherry, Horch, Isaacson,
  Everett, Fischer, Torres, Johnson, Endl, MacQueen, Bryson, Dotson, Haas,
  Kolodziejczak, Van~Cleve, Chandrasekaran, Twicken, Quintana, Clarke, Allen,
  Li, Wu, Tenenbaum, Verner, Bruhweiler, Barnes, \& Prsa}]{Borucki10}
Borucki, W.~J., Koch, D., Basri, G., {et~al.} 2010, Science, 327, 977

\bibitem[{{Brooke} {et~al.}(2014){Brooke}, {Ram}, {Western}, {Li}, {Schwenke},
  \& {Bernath}}]{Brooke14}
{Brooke}, J. S.~A., {Ram}, R.~S., {Western}, C.~M., {et~al.} 2014, \apjs, 210,
  23

\bibitem[{{Bus} {et~al.}(1991){Bus}, {A'Hearn}, {Schleicher}, \&
  {Bowell}}]{Bus1991}
{Bus}, S.~J., {A'Hearn}, M.~F., {Schleicher}, D.~G., \& {Bowell}, E. 1991,
  Science, 251, 774

\bibitem[{{Cochran} \& {Barker}(1986)}]{Cochran86}
{Cochran}, A.~L. \& {Barker}, E.~S. 1986, in ESA Special Publication, Vol. 250,
  ESLAB Symposium on the Exploration of Halley's Comet, ed. B.~{Battrick},
  E.~J. {Rolfe}, \& R.~{Reinhard}, 439

\bibitem[{{Combi} \& {Delsemme}(1980)}]{Combi80}
{Combi}, M.~R. \& {Delsemme}, A.~H. 1980, \apj, 237, 641

\bibitem[{{Cordiner} {et~al.}(2014){Cordiner}, {Remijan}, {Boissier}, {Milam},
  {Mumma}, {Charnley}, {Paganini}, {Villanueva}, {Bockel{\'e}e-Morvan}, {Kuan},
  {Chuang}, {Lis}, {Biver}, {Crovisier}, {Minniti}, \&
  {Coulson}}]{Cordiner2014}
{Cordiner}, M.~A., {Remijan}, A.~J., {Boissier}, J., {et~al.} 2014, \apj, 792,
  L2

\bibitem[{{Cremonese} {et~al.}(1997){Cremonese}, {Boehnhardt}, {Crovisier},
  {Rauer}, {Fitzsimmons}, {Fulle}, {Licandro}, {Pollacco}, {Tozzi}, \&
  {West}}]{Cremonese1997}
{Cremonese}, G., {Boehnhardt}, H., {Crovisier}, J., {et~al.} 1997, \apj, 490,
  L199

\bibitem[{{Cremonese} {et~al.}(2002){Cremonese}, {Huebner}, {Rauer}, \&
  {Boice}}]{Cremonese02}
{Cremonese}, G., {Huebner}, W.~F., {Rauer}, H., \& {Boice}, D.~C. 2002,
  Advances in Space Research, 29, 1187

\bibitem[{{Dufay} {et~al.}(1965){Dufay}, {Swings}, \& {Fehrenbach}}]{Dufay65}
{Dufay}, J., {Swings}, P., \& {Fehrenbach}, C. 1965, \apj, 142, 1698

\bibitem[{{Ferlet} {et~al.}(1987){Ferlet}, {Hobbs}, \& {Madjar}}]{Ferlet87}
{Ferlet}, R., {Hobbs}, L.~M., \& {Madjar}, A.~V. 1987, \aap, 185, 267

\bibitem[{{Fitzsimmons} \& {Cartwright}(1996)}]{Fitzsimmons1996}
{Fitzsimmons}, A. \& {Cartwright}, I.~M. 1996, \mnras, 278, L37

\bibitem[{{Fray} {et~al.}(2005){Fray}, {B{\'e}nilan}, {Cottin}, {Gazeau}, \&
  {Crovisier}}]{Fray2005}
{Fray}, N., {B{\'e}nilan}, Y., {Cottin}, H., {Gazeau}, M.~C., \& {Crovisier},
  J. 2005, \planss, 53, 1243

\bibitem[{{Fulle} {et~al.}(2007){Fulle}, {Leblanc}, {Harrison}, {Davis},
  {Eyles}, {Halain}, {Howard}, {Bockel{\'e}e-Morvan}, {Cremonese}, \&
  {Scarmato}}]{Fulle2007}
{Fulle}, M., {Leblanc}, F., {Harrison}, R.~A., {et~al.} 2007, \apj, 661, L93

\bibitem[{{Fulle} {et~al.}(2013){Fulle}, {Molaro}, {Buzzi}, \&
  {Valisa}}]{Fulle13}
{Fulle}, M., {Molaro}, P., {Buzzi}, L., \& {Valisa}, P. 2013, \apjl, 771, L21

\bibitem[{{Gray} {et~al.}(2006){Gray}, {Corbally}, {Garrison}, {McFadden},
  {Bubar}, {McGahee}, {O'Donoghue}, \& {Knox}}]{Gray06}
{Gray}, R.~O., {Corbally}, C.~J., {Garrison}, R.~F., {et~al.} 2006, \aj, 132,
  161

\bibitem[{{Harris} {et~al.}(2002){Harris}, {Scherb}, {Mierkiewicz},
  {Oliversen}, \& {Morgenthaler}}]{Harris2002}
{Harris}, W.~M., {Scherb}, F., {Mierkiewicz}, E., {Oliversen}, R., \&
  {Morgenthaler}, J. 2002, \apj, 578, 996

\bibitem[{{Haser} {et~al.}(2020){Haser}, {Oset}, \& {Bodewits}}]{Haser2020}
{Haser}, L., {Oset}, S., \& {Bodewits}, D. 2020, Planetary Sciences Journal, 1,
  83

\bibitem[{{Heller}(2024)}]{Heller24}
{Heller}, R. 2024, \aap, 689, A97

\bibitem[{{Hobbs} {et~al.}(1985){Hobbs}, {Vidal-Madjar}, {Ferlet}, {Albert}, \&
  {Gry}}]{Hobbs85}
{Hobbs}, L.~M., {Vidal-Madjar}, A., {Ferlet}, R., {Albert}, C.~E., \& {Gry}, C.
  1985, \apjl, 293, L29

\bibitem[{{Huebner} {et~al.}(1992){Huebner}, {Keady}, \& {Lyon}}]{Huebner92}
{Huebner}, W.~F., {Keady}, J.~J., \& {Lyon}, S.~P. 1992, \apss, 195, 1

\bibitem[{{Hui}(2023)}]{Hui2023}
{Hui}, M.-T. 2023, \aj, 165, 94

\bibitem[{Hunter(2007)}]{hunter2007}
Hunter, J.~D. 2007, Computing In Science \& Engineering, 9, 90

\bibitem[{{Hyland} {et~al.}(2019){Hyland}, {Fitzsimmons}, \&
  {Snodgrass}}]{hyland2019}
{Hyland}, M.~G., {Fitzsimmons}, A., \& {Snodgrass}, C. 2019, \mnras, 484, 1347

\bibitem[{{Keller} {et~al.}(1987){Keller}, {Delamere}, {Reitsema}, {Huebner},
  \& {Schmidt}}]{Keller1987}
{Keller}, H.~U., {Delamere}, W.~A., {Reitsema}, H.~J., {Huebner}, W.~F., \&
  {Schmidt}, H.~U. 1987, \aap, 187, 807

\bibitem[{{Kelley} {et~al.}(2021){Kelley}, {Farnham}, {Li}, {Bodewits},
  {Snodgrass}, {Allen}, {Bellm}, {Coughlin}, {Drake}, {Duev}, {Graham},
  {Kupfer}, {Masci}, {Reiley}, {Walters}, {Dominik}, {J{\o}rgensen}, {Andrews},
  {Bach-M{\o}ller}, {Bozza}, {Burgdorf}, {Campbell-White}, {Dib}, {Fujii},
  {Hinse}, {Hundertmark}, {Khalouei}, {Longa-Pe{\~n}a}, {Rabus}, {Rahvar},
  {Sajadian}, {Skottfelt}, {Southworth}, {Tregloan-Reed}, {Unda-Sanzana}, \&
  {Mindstep Collaboration}}]{Kelley2021}
{Kelley}, M. S.~P., {Farnham}, T.~L., {Li}, J.-Y., {et~al.} 2021, Planetary
  Sciences Journal, 2, 131

\bibitem[{{Kennedy}(2018)}]{Kennedy18}
{Kennedy}, G.~M. 2018, \mnras, 479, 1997

\bibitem[{{Kennedy} {et~al.}(2019){Kennedy}, {Hope}, {Hodgkin}, \&
  {Wyatt}}]{Kennedy19}
{Kennedy}, G.~M., {Hope}, G., {Hodgkin}, S.~T., \& {Wyatt}, M.~C. 2019, \mnras,
  482, 5587

\bibitem[{{Kiefer} {et~al.}(2014){Kiefer}, {Lecavelier des Etangs}, {Boissier},
  {Vidal-Madjar}, {Beust}, {Lagrange}, {H{\'e}brard}, \& {Ferlet}}]{Kiefer14}
{Kiefer}, F., {Lecavelier des Etangs}, A., {Boissier}, J., {et~al.} 2014, \nat,
  514, 462

\bibitem[{{Kiefer} {et~al.}(2017){Kiefer}, {Lecavelier des {\'E}tangs},
  {Vidal-Madjar}, {H{\'e}brard}, {Bourrier}, \& {Wilson}}]{Kiefer17}
{Kiefer}, F., {Lecavelier des {\'E}tangs}, A., {Vidal-Madjar}, A., {et~al.}
  2017, \aap, 608, A132

\bibitem[{{Kiefer} {et~al.}(2019){Kiefer}, {Vidal-Madjar}, {Lecavelier des
  Etangs}, {Bourrier}, {Ehrenreich}, {Ferlet}, {H{\'e}brard}, \&
  {Wilson}}]{Kiefer2018}
{Kiefer}, F., {Vidal-Madjar}, A., {Lecavelier des Etangs}, A., {et~al.} 2019,
  \aap, 621, A58

\bibitem[{{Koen} {et~al.}(2003){Koen}, {Balona}, {Khadaroo}, {Lane},
  {Prinsloo}, {Smith}, \& {Laney}}]{Koen03}
{Koen}, C., {Balona}, L.~A., {Khadaroo}, K., {et~al.} 2003, \mnras, 344, 1250

\bibitem[{{Lagrange-Henri} {et~al.}(1992){Lagrange-Henri}, {Gosset}, {Beust},
  {Ferlet}, \& {Vidal-Madjar}}]{LagrangeHenri92}
{Lagrange-Henri}, A.~M., {Gosset}, E., {Beust}, H., {Ferlet}, R., \&
  {Vidal-Madjar}, A. 1992, \aap, 264, 637

\bibitem[{{Lecavelier des Etangs} {et~al.}(2022){Lecavelier des Etangs},
  {Cros}, {H{\'e}brard}, {Martioli}, {Duquesnoy}, {Kenworthy}, {Kiefer},
  {Lacour}, {Lagrange}, {Meunier}, \& {Vidal-Madjar}}]{LecavelierdesEtangs22}
{Lecavelier des Etangs}, A., {Cros}, L., {H{\'e}brard}, G., {et~al.} 2022,
  Scientific Reports, 12, 5855

\bibitem[{{Lecavelier Des Etangs} {et~al.}(1999){Lecavelier Des Etangs},
  {Vidal-Madjar}, \& {Ferlet}}]{1999A&A...343..916L}
{Lecavelier Des Etangs}, A., {Vidal-Madjar}, A., \& {Ferlet}, R. 1999, \aap,
  343, 916

\bibitem[{{Lee} {et~al.}(2024){Lee}, {Gaidos}, {van Saders}, {Feiden}, \&
  {Gagn{\'e}}}]{Lee24}
{Lee}, R.~A., {Gaidos}, E., {van Saders}, J., {Feiden}, G.~A., \& {Gagn{\'e}},
  J. 2024, \mnras, 528, 4760

\bibitem[{{Lindegren} {et~al.}(2021){Lindegren}, {Klioner}, {Hern{\'a}ndez},
  {Bombrun}, {Ramos-Lerate}, {Steidelm{\"u}ller}, {Bastian}, {Biermann}, {de
  Torres}, {Gerlach}, {Geyer}, {Hilger}, {Hobbs}, {Lammers}, {McMillan},
  {Stephenson}, {Casta{\~n}eda}, {Davidson}, {Fabricius}, {Gracia-Abril},
  {Portell}, {Rowell}, {Teyssier}, {Torra}, {Bartolom{\'e}}, {Clotet},
  {Garralda}, {Gonz{\'a}lez-Vidal}, {Torra}, {Abbas}, {Altmann}, {Anglada
  Varela}, {Balaguer-N{\'u}{\~n}ez}, {Balog}, {Barache}, {Becciani}, {Bernet},
  {Bertone}, {Bianchi}, {Bouquillon}, {Brown}, {Bucciarelli}, {Busonero},
  {Butkevich}, {Buzzi}, {Cancelliere}, {Carlucci}, {Charlot}, {Cioni},
  {Crosta}, {Crowley}, {del Peloso}, {del Pozo}, {Drimmel}, {Esquej}, {Fienga},
  {Fraile}, {Gai}, {Garcia-Reinaldos}, {Guerra}, {Hambly}, {Hauser},
  {Jan{\ss}en}, {Jordan}, {Kostrzewa-Rutkowska}, {Lattanzi}, {Liao}, {Licata},
  {Lister}, {L{\"o}ffler}, {Marchant}, {Masip}, {Mignard}, {Mints}, {Molina},
  {Mora}, {Morbidelli}, {Murphy}, {Pagani}, {Panuzzo}, {Pe{\~n}alosa Esteller},
  {Poggio}, {Re Fiorentin}, {Riva}, {Sagrist{\`a} Sell{\'e}s}, {Sanchez
  Gimenez}, {Sarasso}, {Sciacca}, {Siddiqui}, {Smart}, {Souami}, {Spagna},
  {Steele}, {Taris}, {Utrilla}, {van Reeven}, \& {Vecchiato}}]{Lindegren21}
{Lindegren}, L., {Klioner}, S.~A., {Hern{\'a}ndez}, J., {et~al.} 2021, \aap,
  649, A2

\bibitem[{{Luger} {et~al.}(2021){Luger}, {Bedell}, {Foreman-Mackey},
  {Crossfield}, {Zhao}, \& {Hogg}}]{Luger2021}
{Luger}, R., {Bedell}, M., {Foreman-Mackey}, D., {et~al.} 2021, arXiv e-prints,
  arXiv:2110.06271

\bibitem[{Müller {et~al.}(2024)Müller, Altwegg, Berthelier, Combi,
  De Keyser, Fuselier, Garnier, Hänni, Mall, Rubin, Wampfler, \&
  Wurz}]{Muller2024}
Müller, D.~R., Altwegg, K., Berthelier, J.-J., {et~al.} 2024, Monthly Notices
  of the Royal Astronomical Society, 529, 2763

\bibitem[{{O'Brien} {et~al.}(2018){O'Brien}, {Izidoro}, {Jacobson}, {Raymond},
  \& {Rubie}}]{Obrien2018}
{O'Brien}, D.~P., {Izidoro}, A., {Jacobson}, S.~A., {Raymond}, S.~N., \&
  {Rubie}, D.~C. 2018, \ssr, 214, 47

\bibitem[{{Pavlenko} {et~al.}(2022){Pavlenko}, {Kulyk}, {Shubina}, {Vasylenko},
  {Dobrycheva}, \& {Korsun}}]{Pavlenko22}
{Pavlenko}, Y., {Kulyk}, I., {Shubina}, O., {et~al.} 2022, \aap, 660, A49

\bibitem[{{Pollard}(2024)}]{Pollard24}
{Pollard}, K. 2024, in AAS/Division for Extreme Solar Systems Abstracts,
  Vol.~56, AAS/Division for Extreme Solar Systems Abstracts, 621.12

\bibitem[{{Preston}(1967)}]{Preston1967}
{Preston}, G.~W. 1967, \apj, 147, 718

\bibitem[{{Rappaport} {et~al.}(2018){Rappaport}, {Vanderburg}, {Jacobs},
  {LaCourse}, {Jenkins}, {Kraus}, {Rizzuto}, {Latham}, {Bieryla}, {Lazarevic},
  \& {Schmitt}}]{Rappaport18}
{Rappaport}, S., {Vanderburg}, A., {Jacobs}, T., {et~al.} 2018, \mnras, 474,
  1453

\bibitem[{Rossum(1995)}]{rossum1995}
Rossum, G. 1995, Python Reference Manual, Tech. rep., Centrum voor Wiskunde en
  Informatica (CWI), Amsterdam, The Netherlands, The Netherlands

\bibitem[{{Schleicher}(2010)}]{Schleicher2010}
{Schleicher}, D.~G. 2010, \aj, 140, 973

\bibitem[{{Schleicher} {et~al.}(2024){Schleicher}, {Birch}, {Farnham}, \&
  {Bair}}]{Schleicher2024}
{Schleicher}, D.~G., {Birch}, P.~V., {Farnham}, T.~L., \& {Bair}, A.~N. 2024,
  Planetary Sciences Journal, 5, 281

\bibitem[{{Schmidt}(2016)}]{Schmidt16}
{Schmidt}, C. 2016, \icarus, 265, 35

\bibitem[{{Slaughter}(1969)}]{Slaughter1969}
{Slaughter}, C.~D. 1969, \aj, 74, 929

\bibitem[{{Sunshine} \& {Feaga}(2021)}]{Sunshine2021}
{Sunshine}, J.~M. \& {Feaga}, L.~M. 2021, Planetary Sciences Journal, 2, 92

\bibitem[{{Tennyson} \& {Yurchenko}(2012)}]{Tennyson12}
{Tennyson}, J. \& {Yurchenko}, S.~N. 2012, \mnras, 425, 21

\bibitem[{Tennyson {et~al.}(2016)Tennyson, Yurchenko, Al-Refaie, Barton, Chubb,
  Coles, Diamantopoulou, Gorman, Hill, Lam, Lodi, McKemmish, Na, Owens,
  Polyansky, Rivlin, Sousa-Silva, Underwood, Yachmenev, \& Zak}]{Tennyson16}
Tennyson, J., Yurchenko, S.~N., Al-Refaie, A.~F., {et~al.} 2016, Journal of
  Molecular Spectroscopy, 327, 73, new Visions of Spectroscopic Databases,
  Volume II

\bibitem[{{Thackeray} {et~al.}(1966){Thackeray}, {Feast}, \&
  {Warner}}]{Thackeray1966}
{Thackeray}, A.~D., {Feast}, M.~W., \& {Warner}, B. 1966, \apj, 143, 276

\bibitem[{{Tobin} {et~al.}(2019){Tobin}, {Barnes}, {Persson}, \&
  {Pollard}}]{Tobin19}
{Tobin}, W., {Barnes}, S.~I., {Persson}, S., \& {Pollard}, K.~R. 2019, \mnras,
  489, 574

\bibitem[{van~der Walt {et~al.}(2011)van~der Walt, Colbert, \&
  Varoquaux}]{walt2011}
van~der Walt, S., Colbert, S.~C., \& Varoquaux, G. 2011, Computing in Science
  \& Engineering, 13, 22

\bibitem[{{Vidal-Madjar} {et~al.}(1986){Vidal-Madjar}, {Hobbs}, {Ferlet},
  {Gry}, \& {Albert}}]{VidalMadjar86}
{Vidal-Madjar}, A., {Hobbs}, L.~M., {Ferlet}, R., {Gry}, C., \& {Albert}, C.~E.
  1986, \aap, 167, 325

\bibitem[{Virtanen {et~al.}(2020)Virtanen, Gommers, Oliphant, Haberland, Reddy,
  Cournapeau, Burovski, Peterson, Weckesser, Bright, {van der Walt}, Brett,
  Wilson, Millman, Mayorov, Nelson, Jones, Kern, Larson, Carey, Polat, Feng,
  Moore, {VanderPlas}, Laxalde, Perktold, Cimrman, Henriksen, Quintero, Harris,
  Archibald, Ribeiro, Pedregosa, {van Mulbregt}, \& {SciPy 1.0
  Contributors}}]{2020SciPy-NMeth}
Virtanen, P., Gommers, R., Oliphant, T.~E., {et~al.} 2020, Nature Methods, 17,
  261

\bibitem[{{Weaver} {et~al.}(1997){Weaver}, {Feldman}, {A'Hearn}, \&
  {Arpigny}}]{Weaver1997}
{Weaver}, H.~A., {Feldman}, P.~D., {A'Hearn}, M.~F., \& {Arpigny}, C. 1997,
  Science, 275, 1900

\bibitem[{{Wilson} {et~al.}(2017){Wilson}, {Lecavelier des Etangs},
  {Vidal-Madjar}, {Bourrier}, {H{\'e}brard}, {Kiefer}, {Beust}, {Ferlet}, \&
  {Lagrange}}]{Wilson17}
{Wilson}, P.~A., {Lecavelier des Etangs}, A., {Vidal-Madjar}, A., {et~al.}
  2017, \aap, 599, A75

\bibitem[{{Woods} {et~al.}(2000){Woods}, {Tobiska}, {Rottman}, \&
  {Worden}}]{Woods00}
{Woods}, T.~N., {Tobiska}, W.~K., {Rottman}, G.~J., \& {Worden}, J.~R. 2000,
  \jgr, 105, 27195

\bibitem[{{Zieba} {et~al.}(2019){Zieba}, {Zwintz}, {Kenworthy}, \&
  {Kennedy}}]{Zieba19}
{Zieba}, S., {Zwintz}, K., {Kenworthy}, M.~A., \& {Kennedy}, G.~M. 2019, \aap,
  625, L13

\end{thebibliography}

\clearpage
\FloatBarrier

\begin{appendix}
    
\section{Selected exocomets}\label{sec:app_FEBs}
The dates for the nights selected to have the strongest exocomet features, as well as the depths and velocities of these features can be found in Table~\ref{tab:FEB_data}.
\begin{table}[h]
\caption{Nights with the strongest exocomet features.}
\label{tab:FEB_data}
\begin{tabular}{c c c}
   BJD-2,400,000\tablefootmark{a}  &    v$_{comet}$  &  Depth\\
 & [km/s] & \\
 \hline
\hline
53031.2 & 19.7 & 0.63\\
53032.2 & 18.9 & 0.74\\
53033.1 & 22.3 & 0.76\\
53036.1 & 15.5 & 0.67\\
53045.1 & 18.0 & 0.72\\
53046.1 & 18.0 & 0.91\\
53047.1 & 20.6 & 0.84\\
53048.0 & 20.6 & 0.83\\
53049.1 & 20.6 & 0.94\\
53050.1 & 18.9 & 0.90\\
53051.1 & 18.9 & 0.87\\
53052.1 & 18.9 & 0.92\\
53053.1 & 18.0 & 0.93\\
53054.1 & 17.2 & 0.83\\
53056.1 & 16.3 & 0.86\\
53057.1 & 15.5 & 0.93\\
53058.1 & 14.6 & 0.92\\
53059.1 & 14.6 & 0.87\\
53060.1 & 13.7 & 0.82\\
53061.1 & 13.7 & 0.72\\
53062.1 & 12.9 & 0.69\\
53063.1 & 12.0 & 0.67\\
53064.1 & 12.0 & 0.61\\
53581.4 & -8.5 & 0.65\\
54223.0 &  8.6 & 0.62\\
54542.1 & 29.2 & 0.91\\
54544.1 & 28.3 & 0.92\\
54545.0 & 23.2 & 0.79\\
54546.0 &  8.6 & 0.80\\
54799.2 &  8.6 & 0.75\\
54878.1 & 14.6 & 0.76\\
54913.0 & 12.0 & 0.81\\
56524.4 & 31.7 & 0.76\\
56534.3 & 12.9 & 0.76\\
56536.3 &  8.6 & 0.61\\
57849.0 & 10.3 & 0.73\\
\hline
\end{tabular}
\tablefoot{\tablefoottext{a}{Dates given are for the average of all the observations that were combined during each of the nights.}}
\end{table}

\end{appendix}

\end{document}